\documentclass[10pt,aps,prd,reprint,nofootinbib,floats,floatfix,amsfonts,amssymb,amsmath,preprintnumbers,notitlepage,superscriptaddress]{revtex4-1}
\usepackage{aastex_macros}
\usepackage{float}
\usepackage{graphicx}
\usepackage{caption}
\usepackage{subcaption}
\usepackage[utf8]{inputenc}
\usepackage[british]{babel}
\usepackage{isomath}
\usepackage[protrusion=true,tracking=true,kerning=true,spacing=true,final,babel=true]{microtype}
\usepackage{physics}
\usepackage{amsmath}
\usepackage{xcolor}
\usepackage{url}
\usepackage[final]{hyperref}

\begin{document}

\title{Searching for parity violation with the LIGO-Virgo-KAGRA network}
\author{Katarina Martinovic}
\affiliation{Theoretical Particle Physics and Cosmology Group, \, Physics \, Department, \\ King's College London, \, University \, of London, \, Strand, \, London \, WC2R \, 2LS, \, UK}
\author{Charles Badger}
\affiliation{Theoretical Particle Physics and Cosmology Group, \, Physics \, Department, \\ King's College London, \, University \, of London, \, Strand, \, London \, WC2R \, 2LS, \, UK}
\author{Mairi Sakellariadou}
\affiliation{Theoretical Particle Physics and Cosmology Group, \, Physics \, Department, \\ King's College London, \, University \, of London, \, Strand, \, London \, WC2R \, 2LS, \, UK}
\affiliation{Theoretical Physics Department, CERN, Geneva, Switzerland}
\author{Vuk Mandic}
\affiliation{School of Physics and Astronomy,
University of Minnesota, Minneapolis, MN 55455 USA}
\date{\today}

\begin{abstract}
A stochastic gravitational wave background is expected to emerge from the superposition of numerous gravitational wave sources of both astrophysical and cosmological origin. A number of cosmological models can have a parity violation, resulting in the generation of circularly polarised gravitational waves. We present a method to search for parity violation in the gravitational wave data. We first apply this method to the most recent, third, LIGO-Virgo observing run. We then investigate the constraining power of future A+ LIGO-Virgo detectors, including KAGRA to the network, for a gravitational wave background generated by early universe cosmological turbulence. 
\end{abstract}

\maketitle

\section{Introduction}

A stochastic gravitational wave background (SGWB) is expected to be created from the overlap of gravitational waves (GWs) coming from many independent sources. A number of early universe cosmological sources have been proposed, including GWs sourced from inflation \cite{AxInf_GW}, cosmic strings \cite{CosStr_GW}, first order phase  transitions (for recent reviews see, e.g. \cite{Caprini:2015zlo,Hindmarsh:2020hop}), or cosmological models inspired from string theory (see, e.g. \cite{altCos_GW,altCos2_GW}). Searches for an unpolarised isotropic SGWB have been conducted in the past using data gathered by ground based interferometer detectors LIGO \cite{TheLIGOScientific:2014jea} and Virgo \cite{TheVirgo:2014hva}, and allowed upper limits on SGWB energy density to be placed \cite{LIGOScientific:2019vic, Abbott:2021xxi, O3Data}. 

A number of mechanisms in the early universe can create parity violation \cite{Alexander_2006} that may manifest itself in the production of asymmetric amounts of right- and left-handed circularly polarised isotropic GWs. Since astrophysical sources of the SGWB are unlikely to have this circular polarisation, a detection of such a signal can allow cosmologically sourced GWs to be distinguished from the astrophysically sourced component of the SGWB. A closer analysis of polarised SGWB can place constraints on parity violating theories.

Numerous parity violating effects on the SGWB have been studied in the literature, including those resulting from the Chern-Simons gravitational term \cite{CS_PV} and axion inflation \cite{AxInf_PV}. Another potential chiral source for early universe SGWB is turbulence in the primordial plasma induced either from cosmological first-order (electroweak or QCD) phase transitions \cite{Kamionkowski_1994,EdWitten_PhaseTrans,Hogan_PhaseTrans}, or from the primordial magnetic fields that are coupled to the cosmological plasma \cite{Brandenburg_1996,Christensson_2001,Kahniashvili_2010,Brandenburg_2019,Brandenburg:2021aln}. Parity violating effects on the SGWB have been explored in detail before ~\cite{Crowder_2013} from a previous LIGO-Virgo observing run ~\cite{obsRun_2009}.

Since turbulence is a stochastic process, the GWs produced in the process are stochastic as well. Similarly, a parity violating turbulent source will produce circularly polarised GWs. Depending on the helicity strength of the source, there are two types of turbulence GW spectra \cite{HT_Study,HK_Study}. Turbulence dominated by energy dissipation at small scales leads to a helical Kolmogorov (HK) spectrum, whereas turbulence dominated by helicity dissipation at small scales leads to a helicity transfer (HT) spectrum. We focus on models that result in a HK spectrum, and consider the polarisation degree associated with them. 

In what follows we adopt the formalism of \cite{Formalism_MathMod} and present a method to detect parity violation in GW data. We first analyse recent data from the third Advanced LIGO-Virgo observing (O3) run \cite{Abbott:2021xxi} to place upper limits for a simple power law parity violation model for the normalised GW energy density $\Omega_{\rm GW}$. We consequently study the SGWB produced by turbulence in the primordial plasma  and investigate what upper limits can be placed with the inclusion of KAGRA~\cite{akutsu2020overview} and improved LIGO-Virgo sensitivities.

The rest of the paper is organised as follows:
in Sec.~\ref{sec: methods} we present our methodology which we then apply to some parity violation models in Sec.~\ref{sec: models}. In Sec.~\ref{sec: results} we state our results and in Sec.~\ref{sec: concl} we discuss our conclusions.
 
\section{Method}
\label{sec: methods}

We use the plane-wave expansion of the metric $h_{ab}$ at cosmic time $t$ and position vector $\Vec{x}$~\cite{Romano_Cornish}:
\begin{align}
    h_{ab}(t,\Vec{x}) = \sum_A \int_{-\infty}^{\infty}df \int_{S^2} d\hat{\Omega} h_A(f,\hat{\Omega}) e^{-2\pi if(t-\Vec{x}\cdot\hat{\Omega})} e_{ab}^A(\hat{\Omega}),
    \label{eq:planewave_exp}
\end{align}
 where $f$ is the frequency and $e_{ab}^A(\hat{\Omega})$ 
 is the polarisation tensor for a wave traveling in direction $\hat{\Omega}$. We use the circularly polarised bases $e^R = (e^{+} + ie^{\cross})/\sqrt{2}$ and $e^L = (e^{+} - ie^{\cross})/\sqrt{2}$ (with + and $\cross$ the plus and cross polarisations, respectively)  to obtain the right- and left-handed modes $h_R = (h_{+} - ih_{\cross})/\sqrt{2}$ and $h^L = (h_{+} + ih_{\cross})/\sqrt{2}$, respectively. Right- and left-handed correlators can then be written as
\begin{multline}
    \begin{pmatrix}
    \langle h_{R}(f,\hat{\Omega})h_{R}^{*}(f',\hat{\Omega}') \rangle \\
    \langle h_{L}(f,\hat{\Omega})h_{L}^{*}(f',\hat{\Omega}') \rangle
    \end{pmatrix} \\
    = \frac{\delta(f-f')\delta^2(\hat{\Omega}-\hat{\Omega}')}{4\pi}
    \begin{pmatrix}
    I(f,\hat{\Omega}) + V(f,\hat{\Omega}) \\
    I(f,\hat{\Omega}) - V(f,\hat{\Omega})
    \end{pmatrix},
    \label{eq:RL_correlators}
\end{multline}
where $\langle\cdot\rangle$ represents the ensemble average and $I, V$ are the Stokes parameters, with $V$ characterising the asymmetry between right- and left-handed polarised waves, and $I (\geq |V|)$ the wave's total amplitude. For $V=0$,  Eq.~(\ref{eq:RL_correlators}) would be simply the correlator for unpolarised isotropic SGWB.

We use the standard cross-correlation estimator ~\cite{Romano_Cornish}:
\begin{eqnarray}
    \langle \hat{C}_{d_1 d_2} \rangle &=&\int_{-\infty}^{\infty}df\int_{-\infty}^{\infty}df'\delta_T(f-f')\langle s_{d_1}^{*}(f)s_{d_2}(f') \rangle \Tilde{Q}(f')\nonumber \\
    &=& \frac{3H_0^2 T}{10\pi^2}\int_0^{\infty}df\frac{\Omega'_{\rm GW}(f)\gamma_I^{d_1 d_2}(f)\Tilde{Q}(f)}{f^3},
    \label{eq:Y_estimator}
\end{eqnarray}
where
\begin{eqnarray}
    \Omega'_{\rm GW} &=& \Omega_{\rm GW}\bigg[1+\Pi(f)\frac{\gamma_V^{d_1 d_2}(f)}{\gamma_I^{d_1 d_2}(f)}\bigg],\nonumber \\
    \gamma_I^{d_1 d_2}(f) &=& \frac{5}{8\pi}\int d\hat{\Omega}(F_{d_1}^{+}F_{d_2}^{+*} + F_{d_1}^{\cross}F_{d_2}^{\cross*})e^{2\pi if\hat{\Omega}\cdot\Delta\Vec{x}}, \nonumber\\
    \gamma_V^{d_1 d_2}(f) &=& -\frac{5}{8\pi}\int d\hat{\Omega}(F_{d_1}^{+}F_{d_2}^{\cross*} - F_{d_1}^{\cross}F_{d_2}^{+*})e^{2\pi if\hat{\Omega}\cdot\Delta\Vec{x}},
    \label{eq:PVOmeg&ORF}
\end{eqnarray}
with $H_0$ the Hubble parameter, $T$ the measurement time, $\delta_T(f) = \sin(\pi fT)/(\pi f)$, 
$\Tilde{s}_{d_1}(f)$ and $\Tilde{s}_{d_2}(f)$ the Fourier transforms of the strain time series of two GW detectors (denoted by $d_1, d_2$). $\Tilde{Q}(f)$ is a filter and $F_n^A = e_{ab}^A d_n^{ab}$ stands for the contraction of the tensor modes of polarisation $A$ to the $n^{\rm th}$ detector's geometry. We denote by 
$\gamma_I^{d_1 d_2}$ 
the standard overlap reduction function of two detectors $d_1, d_2$, and by 
$\gamma_V^{d_1 d_2}$ 
the overlap function associated with the parity violation term. The polarisation degree, $\Pi(f) = V(f)/I(f)$, takes on values between -1 (fully left polarisation) and 1 (fully right polarisation), with $\Pi = 0$ being an unpolarised isotropic SGWB.

The variance associated with the estimator $\hat{C}_{d_1 d_2}$ is \cite{Romano_Cornish}
\begin{align}
    \sigma_{d_1 d_2}^2 = \frac{T}{4}\int_0^{\infty}df P_{d_1}(f)P_{d_2}(f)|\Tilde{Q}(f)|^2
    \label{eq:sig_estimator},
\end{align}
where 
$P_{d_1,d_2}(f)$
are the one-sided noise power spectral densities of GW detectors 
$d_1, d_2$.

To proceed we perform parameter estimation and fit GW models to data using a hybrid frequentist-Bayesian approach \cite{Matas:2020roi}. We construct a Gaussian log-likelihood for a multi-baseline network
\begin{align}
    \log p(\hat C(f) | \boldsymbol{\theta})
    \propto \sum_{d_1 d_2 d_3}\sum_{f}\frac{\left[\hat C_{d_1 d_2}(f) - \Omega'_{\rm GW}(f, \boldsymbol{\theta}) \right]^2}{\sigma_{d_1 d_2}^2(f)},
    \label{eq:pe_ms:likelihood}
\end{align}
where 
$\hat{C}_{d_1 d_2}(f)$
is the frequency-dependent cross-correlation estimator of the SGWB  calculated using data from detectors $d_1, d_2$, and 
$\sigma^2_{d_1 d_2}(f)$ 
is its variance ~\cite{Formalism_Source}. The cross-correlation statistics are constructed using strain data from the individual GW detectors. We assume that correlated-noise sources have been either filtered out \cite{Cirone:2018guh} or accounted for \cite{Meyers:2020qrb}. The  normalised GW energy density model we fit to the data is $\Omega'_{\rm GW}(f,  {\boldsymbol \theta})$, with parameters $\boldsymbol\theta$ including both GW parameters as well as parameters of the $\Pi(f)$ model.

Plotting the $\Omega_{\alpha}$ versus $\alpha$ confidence curve generated from O3 data, one observes that it is easier to constrain $\Omega_{\alpha}$ for entirely right-handed polarised GWs ($\Pi = 1$) than it is for  left-handed ($\Pi = -1$) ones.
Excluding the HL detector baseline however, results to a less obvious polarisation bias for right- or left-handed GWs.

\begin{figure}
    \centering
    \includegraphics[width=0.4\textwidth]{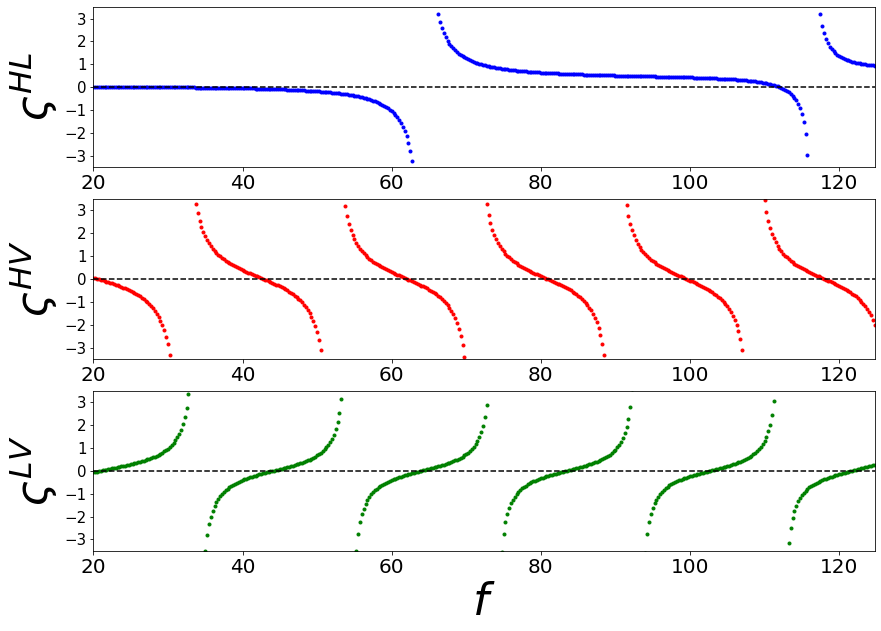}
    \caption{Overlap reduction function (ORF) ratio $\varsigma$ for  HL (top), HV (middle) and LV (bottom) baselines.}
    \label{fig:orf_ratio}
\end{figure}

To understand the origin of this bias we investigate the asymmetry in the overlap reduction ratio, 
$\varsigma^{\rm HL}\equiv\gamma_V^{\rm HL}/\gamma_I^{\rm HL}$,
for the HL detector pairing 
plotted in Fig.~\ref{fig:orf_ratio}, along with the corresponding overlap reduction ratios
$\varsigma^{\rm HV}$
and $\varsigma^{\rm LV}$,
for the HV and LV baselines, respectively.
 We remark that only $\varsigma^{HV}$ and $\varsigma^{LV}$ are roughly periodic in the considered frequency range ($f \lesssim 130 \rm Hz$), whereas $\varsigma^{HL}$ shows different behaviour within this frequency range. It is indeed this non-periodic behavior in $\varsigma^{HL}$ that leads to the polarisation bias and better constraints on right-hand polarised signals.


\section{Models}
\label{sec: models}

We will apply our method in two classes of models.
First, we perform a generic search for a parity violation GW signal, with a power-law behaviour $\Omega_{\rm GW}(f) = \Omega_{\rm ref}\,(f/f_{\rm ref})^{\alpha}$, setting $f_{\rm ref} = 25$ Hz. The amplitude prior we use is log-uniform between $10^{-13}$ and $10^{-5}$, and the spectral index prior is a Gaussian distribution centred at 0 with a standard deviation 3.5, following the priors used in \cite{Abbott:2021xxi}.

We then use a broken power-law spectral shape, motivated by high energy physics.
Extensions of the Standard Model  of particle physics can imply parity violation at the electroweak energy scale being potentially manifested through helical (or chiral) turbulent motion \cite{Long_2014, Dorsch_2017}. Parity violation turbulent sources will produce circularly polarised GWs \cite{Tina_2005}, with a broken power-law spectral shape, peaking at the characteristic frequency of the source. 
The infrared behaviour captured by the rising power-law gives a robust spectral index fixed by causality equal to 3. Recent numerical simulations however, show the flattening of the low-frequency tail \cite{Pol:2019yex} from cubic to linear dependence on frequency. It takes less time for the slope to flatten than it does for the GW signal to become stationary, so we approximate the infrared behaviour as $\propto f$. The decaying power-law at higher frequencies depends on the turbulence model \cite{Weir_2018}.
As an example, in our analysis we focus on
\cite{Tina_2002}:
%
%
\[ \Omega_{\text{GW}}(f) =
  \begin{cases}
    \Omega_{\text{peak}}(f/f_{\rm peak})  &, \quad f\leq f_{\rm peak} \\
    \Omega_{\text{peak}}(f/f_{\rm peak})^{-8/3} &, \quad f > f_{\rm peak}~.
  \end{cases}
  \label{eq: omegaGW_turb}
\]
The peak frequency of the spectrum, $f_{\rm peak}$, is directly related to the temperature, $T_*$, at which the first order phase transition occurred. At the electroweak scale, $T_*=100\rm \, GeV$, we expect the GW spectrum to peak in the mHz range, which has given rise to many LISA-focused turbulence studies. For a higher energy scale $T_* \sim 10^8 \, \rm GeV$, one would see a chiral turbulence spectrum that peaks in the LIGO/Virgo range. It is these early universe signatures we search for with the currently operating GW detectors. For our analysis, we fix $f_{\rm peak} = 25\text{ Hz}$.

In addition to modelling the GW spectrum, we need a model for the amount of polarisation of the turbulence spectrum. Previous studies \cite{Tina_2005,kahniashvili2020circular} calculated numerically  the net circular polarisation of GWs for different initial turbulent conditions to get the polarisation degree $\Pi$ over wavenumber $k$, and found frequency-dependent models of $\Pi$. In the following, we will study both the simplified $\Pi= {\rm const.}$ model, as well as a frequency-dependent polarisation model. In the former simplified case, the prior for $\Pi$ is uniform between -1 and 1. For the latter $\Pi(f)=f^{\beta}$ models, the prior on the spectral index $\beta$ we use is uniform between -2 and 2, but we expect $\Pi$ to decay with frequency ~\cite{kahniashvili2020circular}. Note that we assume this model to be right-handed, since studies \cite{Tina_2005} suggest that the induced GWs would have predominantly right-hand polarisation.

\section{Results}
\label{sec: results}

We first present our results using data from the most recent, third, Advanced LIGO-Virgo data observing run. We then discuss detection prospects of model-dependent turbulence spectra as the sensitivity of the interferometers increases.

\subsection{O3 results}

\begin{figure}
    \centering
    \includegraphics[width=0.49\textwidth]{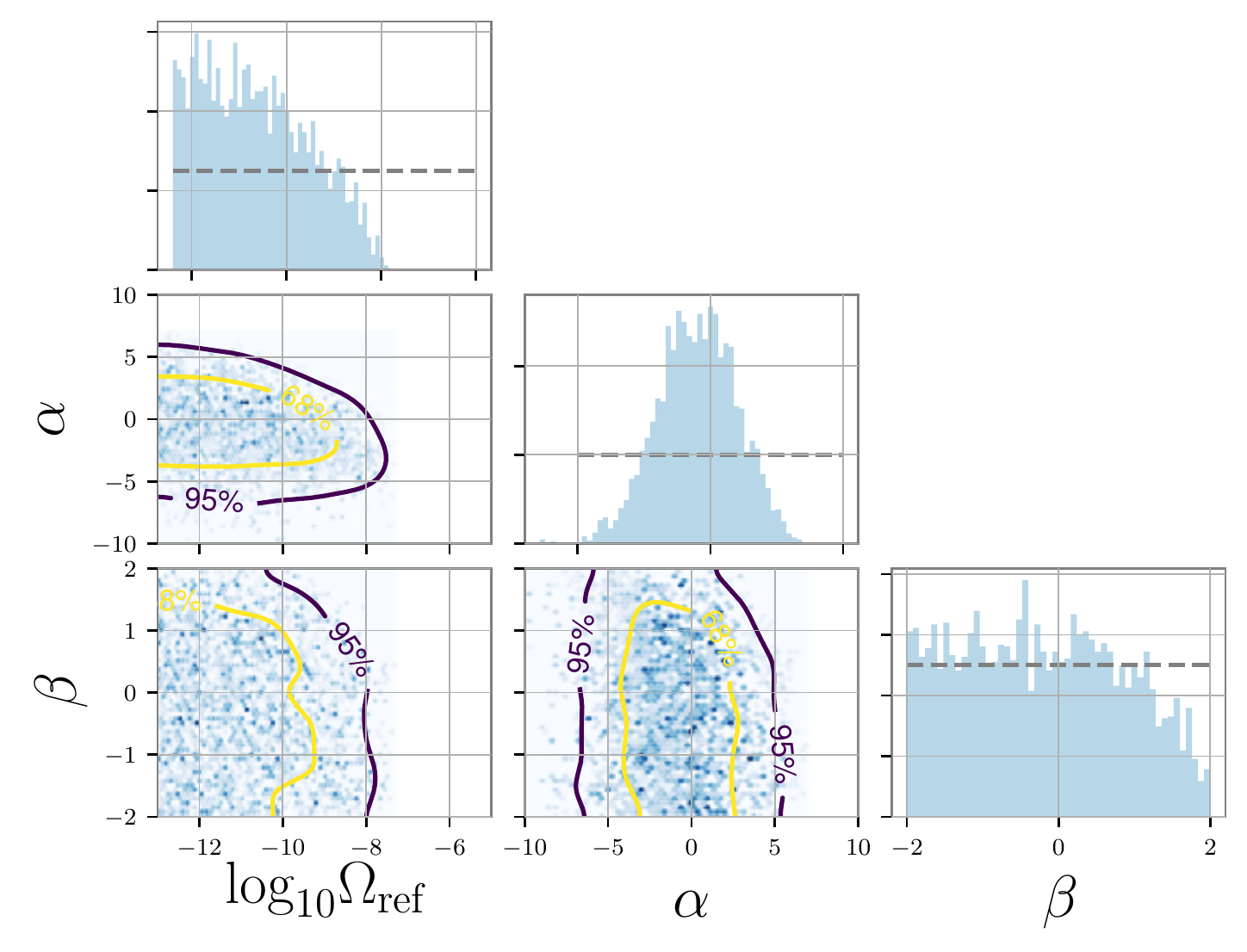}
    \caption{$95\%$ confidence limit and posterior distribution obtained using O3 data for the model of power-law SGWB spectrum with parity violation $\Pi (f) = f^\beta$.}
    \label{fig:O3_beta_corner_plot}
\end{figure}

We search for a parity violation signal in the recent O3 data \cite{O3Data}.  Figure~\ref{fig:O3_beta_corner_plot} shows posterior distribution of amplitude, $\Omega_{\rm GW}$, and spectral index, $\alpha$, of a power-law GW signal, as well as the posterior of the polarisation degree parameter, $\beta$. There is no detection, but we can place an upper limit on the amplitude, $\Omega^{95\%}_{\rm ref}= 4.9 \times 10^{-9}$, after marginalizing over $\alpha$ and $\beta$. The $\alpha$ posterior is similar to the Gaussian prior distribution, implying that we cannot deduce anything about the spectral index, $\alpha$, of the GW power law from the O3 data. Figure \ref{fig:O3_beta_corner_plot} shows that the data favours negative values of $\beta$. This result agrees with previous numerical simulations of turbulent sourced SGWB models \cite{Tina_2005}. 

In addition to the frequency-dependent $\Pi(f)=f^{\beta}$ model, we search for a simpler $\Pi$ = const. polarisation model within the O3 data, to complete our analysis. We find no preference for a particular $\Pi$ value in the [-1,1] prior range. We find the upper limit on the amplitude of the power law to be $\Omega^{95\%}_{\rm ref}= 7.0 \times 10^{-9}$. Calculating the Bayes factor, we find $\rm{ln}\, {\cal{B}}^{\Pi \neq 0}_{\Pi=0} = -0.02$, concluding that there is no preference for parity violation models versus no parity violation ones.

\subsection{Future prospects}

With each observing LIGO-Virgo run, we see improved upper limits on the SGWB, expecting to have a detection in one of the future upgrades of the detector network. Let us therefore investigate the possibility of detecting a parity violation signal with the A+ sensitivity of LIGO, AdV+ sensitivity of Virgo, and including KAGRA at Design sensitivity to the network \cite{Aasi:2013wya}. We simulate the cross-correlation function defined in Eq.~(\ref{eq:Y_estimator}) that contains a GW signal as well as instrumental noise of the detectors. We note that our simulations are for three years observation time. Adding more GW detectors to the network and extending the observation time both lead to improvements in our sensitivity.

We inject a broken power law, $\Omega_{\rm GW}$, induced by turbulence as described in Sec. \ref{sec: models}. To investigate the detection prospects of such a signal, we vary the amplitude of the injected spectrum by doing 1000 injections log-spaced between $\Omega_{\rm peak} \in [10^{-10}, 10^{-7}]$. The polarisation spectrum we inject, $\Pi(f)=f^{-1/2}$, is motivated by recent numerical simulations \cite{kahniashvili2020circular}. We discuss below our results and their dependence on  a deviation of the  polarisation parameter $\beta$ from the $-1/2$ value.

\begin{figure}
    \centering
    \includegraphics[width=0.4\textwidth]{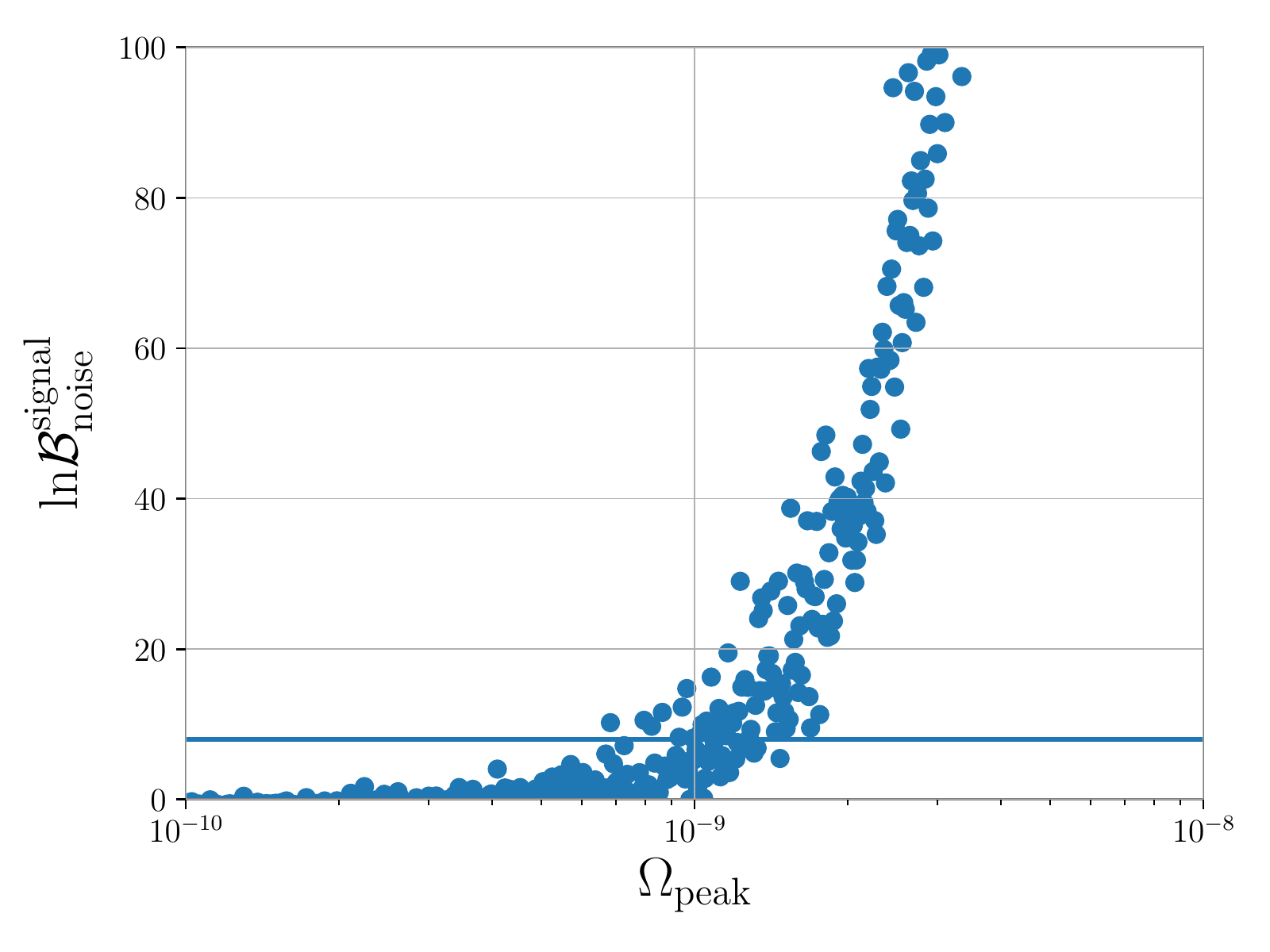}
    \caption{Bayes factor as a function of amplitude of the injected signal. Each point represents one of our 1000 injections. The solid line represents $\rm{ln}\, \cal{B}^{\rm{signal}}_{\rm{noise}}$ = 8.}
    \label{fig:BF}
\end{figure}

Figure~\ref{fig:BF} shows the variation of signal-to-noise Bayes factor, ${\cal B}$, of the injections, focusing particularly in the region $\Omega_{\rm peak} \in [10^{-10}, 10^{-8}]$. A $\ln{\cal B}$ factor of 8 is equivalent to a frequentist SNR of 4 \cite{Romano_Cornish}, and as such, we take this value to be our detection statistic. All injection points that are above the solid line in Fig. \ref{fig:BF} will be detected. This leads to an upper limit of $\Omega_{\rm peak} = 1.5 \times 10^{-9}$; any {\sl louder} signal is expected to be detected with great significance by the A+ detectors. 

\begin{figure}
    \centering
    \includegraphics[width=0.4\textwidth]{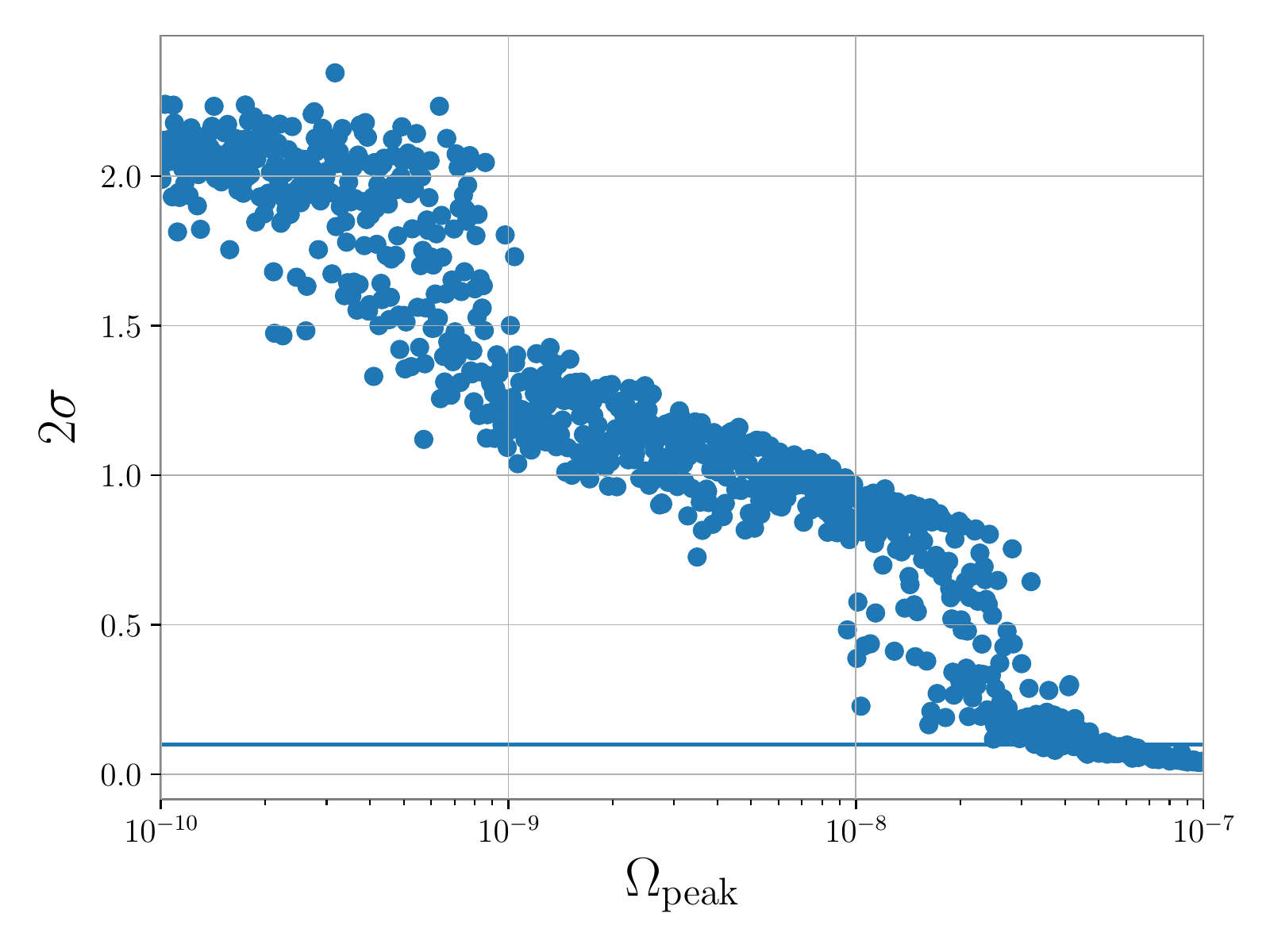}
    \caption{Variation of 2$\sigma$ value of the $\beta$ posterior for each of the 1000 injections. The solid line represents $2\sigma$ = 0.1.}
    \label{fig:sigma}
\end{figure}

However, even if we can confidently claim a detection of a turbulent, broken power-law, GW signal, we may not be able to constrain its polarisation. Repeated injections showed that the strength of the injected signal, $\Omega_{\rm{peak}}$, played a main role in the recovery of the simulated signal. More precisely, we have found that stronger signals yield better results for the recovery of the $\Pi(f)$ model, i.e. the $\beta$ parameter. Our analysis also proved that the inclusion of simulated data from the Virgo and KAGRA detectors was critical in recovering the polarisation of an injected SGWB. We found that for injections with amplitude $\Omega_{\rm peak} \geq 5 \times 10^{-8}$, we confidently recover the $\beta = -1/2$ injected value, see Fig.~\ref{fig:sigma}. We quantify our confidence in recovery of $\beta$ by requiring 95\% (2$\sigma$) of its posterior distribution to be within 0.1 of the injected value.

Therefore, we conclude that the amplitude of the GW spectrum needs to be more than 30 times larger than its detection threshold in order to recover the $\beta=-1/2$ parameter, and detect a polarisation. Only with such a strong detection, one can study the polarisation model and its implications for parity violation theories.

\begin{figure}
    \centering
    \includegraphics[width=0.4\textwidth]{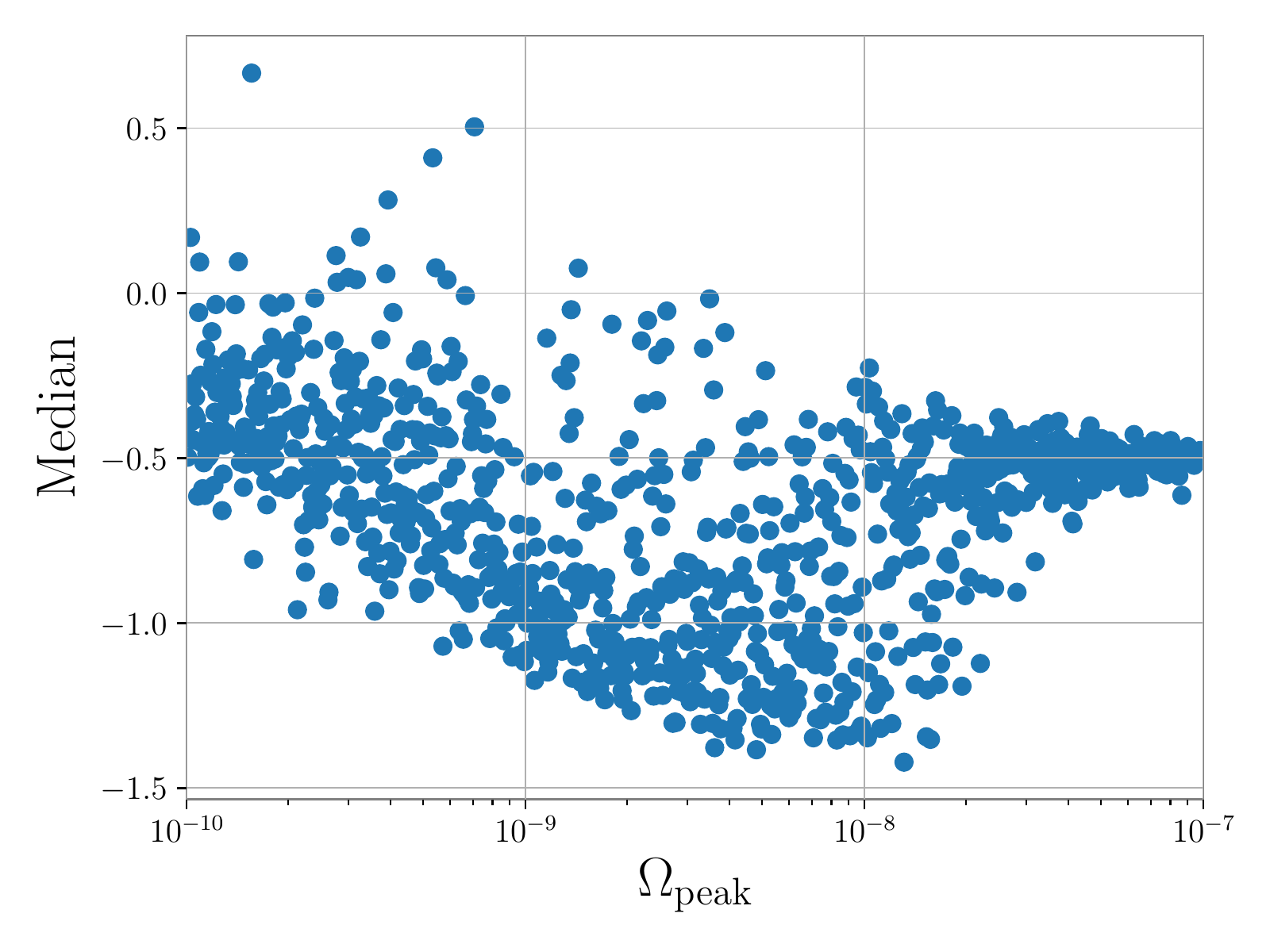}
    \caption{Variation of the median of the $\beta$ posterior for each of the 1000 injections.}
    \label{fig:median}
\end{figure}

Weaker injections will still allow us to place an upper limit on $\beta$, but we will not be able to estimate it. Posteriors for these weak injections will be skewed towards the lower end of the $\beta$ prior. We check median, first and third quartile values of $\beta$ posteriors across the injection range to better understand their skewness. We find that the median of the posterior starts to deviate from $\beta=-1/2$ for $\Omega_{\rm peak} = 5 \times 10^{-8}$, agreeing with our previously stated definition of confident recovery, see Fig. \ref{fig:median}. Finally, the first and third quartile confirm spreading of the posterior towards more negative $\beta$ values.

To check the dependence of our results on the injected value of the $\beta$ parameter, we repeat the analysis for $\beta = -1$ and 0 ~\cite{kahniashvili2020circular}. The detection threshold for each of the data sets is the same as before, $\Omega_{\rm peak} = 1.5 \times 10^{-9}$, suggesting that \textit{when} we claim a detection, it will be independent of the amount of polarisation of the signal. However, the signal strength needed to successfully recover $\beta$ depends on the polarisation model. Namely, the smaller the value of $\beta$ is, the stronger the injection amplitude is needed. For $\beta = -1$ 
we are unable to recover it within our injection range. The $\beta=0$ injection, with a frequency-independent polarisation, is recovered with signals of amplitude $\Omega_{\rm peak} = 1 \times 10^{-8}$,  only 7 times stronger than the injection threshold. The only signal strength that successfully recovers $\beta$, and is \textit{not} constrained by the first three LIGO-Virgo observing runs, is $\beta=0$, implying that even if we include 4 detectors, and consider 3 years of observation time, we cannot probe frequency-dependent models.



%

\section{Conclusions}
\label{sec: concl}

We searched for SGWB generated by parity violation sources in recent GW data (O3) and simulated GW data (future sensitivities of LIGO/Virgo/KAGRA detectors). We found no evidence for such a signal in O3 data, but instead we placed an upper limit on amplitude of a power-law GW model, $\Omega^{95\%}_{\rm ref}= 4.9 \times 10^{-9}$.  Our analysis showed that for a parity violation model $\Pi=f^\beta$, recent O3 data favour models where $\Pi$ decays with frequency, supporting expected numerical simulations ~\cite{kahniashvili2020circular}. A bias for constraining right-handed polarised waves was found to be due to the geometry of the Hanford-Livingston detector baseline, leading to better constraints of $\Pi>0$ polarisations. 

When simulating data for future detection prospects, we considered a chiral turbulence source in the early universe. The results we obtained are model-dependent. For a GW modelled as a broken power-law with lower and higher frequency indices 1 and -8/3, respectively, and peak frequency of 25 Hz, the LIGO-Virgo-KAGRA detector network is sensitive to peak amplitudes down to $\Omega_{\rm peak} = 1.5 \times 10^{-9}$. Our analysis showed that we are able to better estimate the parameters of the parity violation model for stronger simulated GW backgrounds, as well as when the Virgo/KAGRA detectors are included in the analysis, highlighting the importance of having a multi-detector network. For $\beta = -1/2$ we found that for $\beta$ to be successfully recovered, we need injections of at least $\Omega_{\rm peak} = 5 \times 10^{-8}$, which is excluded by existing GW data. Hence, even if we detect a turbulence signal, we may not be able to deduce its polarisation. The recovery of frequency-independent polarisation ($\beta=0$) showed more promising results and we might be able to constrain such signals in the future observing runs.



Although this study used the current LIGO and Virgo detectors as well as the upcoming KAGRA detector, it would be interesting to apply the same study to additional ground detectors added to the network such as LIGO-India, as well as other planned terrestrial detectors (Einstein Telescope ~\cite{Punturo_2010}, Cosmic Explorer ~\cite{reitze2019cosmic}), or the space-detector LISA ~\cite{LISA_Prop}. Note that due to uncertainty in geographical locations (i.e. uncertainty in ORFs) of the planned terrestrial detectors, the study cannot be extended to these at the moment.

In our analysis, we focused on chiral turbulence from an early universe phase transition, but there are other sources of parity violation, like the well-studied chiral inflaton field ~\cite{Sorbo_2011}. The method we have presented here could be easily adapted in such a scenario. 

Unresolved compact binary coalescences (CBCs) are expected to be the dominant contribution to the SGWB. Hence one could study the SGWB with CBC and parity violation signals being both present. 
We leave this as a future work.

\acknowledgments
It is a pleasure to thank Tina Kahniashvili and Axel Brandenburg on fruitful discussions about turbulence. We acknowledge computational resources provided by the LIGO Laboratory and supported by National Science Foundation Grants PHY-0757058 and PHY-0823459. This  paper  has  been given LIGO DCC number LIGO-P2100076.

K.M. is supported by King's College London through a Postgraduate International Scholarship. M.S. is supported in part by the Science and Technology Facility Council (STFC), United Kingdom, under the research grant ST/P000258/1. V.M. is supported by the NSF grant PHY-1806630. 

Software packages used in this paper are \texttt{matplotlib}~\cite{Hunter:2007}, \texttt{numpy}~\cite{numpy}, \texttt{bilby}~\cite{Ashton:2018jfp}, \texttt{ChainConsumer}~\cite{Hinton2016}. 


\bibliography{parity}

\begin{thebibliography}{49}%
\makeatletter
\providecommand \@ifxundefined [1]{%
 \@ifx{#1\undefined}
}%
\providecommand \@ifnum [1]{%
 \ifnum #1\expandafter \@firstoftwo
 \else \expandafter \@secondoftwo
 \fi
}%
\providecommand \@ifx [1]{%
 \ifx #1\expandafter \@firstoftwo
 \else \expandafter \@secondoftwo
 \fi
}%
\providecommand \natexlab [1]{#1}%
\providecommand \enquote  [1]{``#1''}%
\providecommand \bibnamefont  [1]{#1}%
\providecommand \bibfnamefont [1]{#1}%
\providecommand \citenamefont [1]{#1}%
\providecommand \href@noop [0]{\@secondoftwo}%
\providecommand \href [0]{\begingroup \@sanitize@url \@href}%
\providecommand \@href[1]{\@@startlink{#1}\@@href}%
\providecommand \@@href[1]{\endgroup#1\@@endlink}%
\providecommand \@sanitize@url [0]{\catcode `\\12\catcode `\$12\catcode
  `\&12\catcode `\#12\catcode `\^12\catcode `\_12\catcode `\%12\relax}%
\providecommand \@@startlink[1]{}%
\providecommand \@@endlink[0]{}%
\providecommand \url  [0]{\begingroup\@sanitize@url \@url }%
\providecommand \@url [1]{\endgroup\@href {#1}{\urlprefix }}%
\providecommand \urlprefix  [0]{URL }%
\providecommand \Eprint [0]{\href }%
\providecommand \doibase [0]{http://dx.doi.org/}%
\providecommand \selectlanguage [0]{\@gobble}%
\providecommand \bibinfo  [0]{\@secondoftwo}%
\providecommand \bibfield  [0]{\@secondoftwo}%
\providecommand \translation [1]{[#1]}%
\providecommand \BibitemOpen [0]{}%
\providecommand \bibitemStop [0]{}%
\providecommand \bibitemNoStop [0]{.\EOS\space}%
\providecommand \EOS [0]{\spacefactor3000\relax}%
\providecommand \BibitemShut  [1]{\csname bibitem#1\endcsname}%
\let\auto@bib@innerbib\@empty
\bibitem [{\citenamefont {Chung}\ \emph {et~al.}(2000)\citenamefont {Chung},
  \citenamefont {Kolb}, \citenamefont {Riotto},\ and\ \citenamefont
  {Tkachev}}]{AxInf_GW}%
  \BibitemOpen
  \bibfield  {author} {\bibinfo {author} {\bibfnamefont {D.~J.~H.}\
  \bibnamefont {Chung}}, \bibinfo {author} {\bibfnamefont {E.~W.}\ \bibnamefont
  {Kolb}}, \bibinfo {author} {\bibfnamefont {A.}~\bibnamefont {Riotto}}, \ and\
  \bibinfo {author} {\bibfnamefont {I.~I.}\ \bibnamefont {Tkachev}},\ }\href
  {\doibase 10.1103/physrevd.62.043508} {\bibfield  {journal} {\bibinfo
  {journal} {Physical Review D}\ }\textbf {\bibinfo {volume} {62}} (\bibinfo
  {year} {2000}),\ 10.1103/physrevd.62.043508}\BibitemShut {NoStop}%
\bibitem [{\citenamefont {Damour}\ and\ \citenamefont
  {Vilenkin}(2000)}]{CosStr_GW}%
  \BibitemOpen
  \bibfield  {author} {\bibinfo {author} {\bibfnamefont {T.}~\bibnamefont
  {Damour}}\ and\ \bibinfo {author} {\bibfnamefont {A.}~\bibnamefont
  {Vilenkin}},\ }\href {\doibase 10.1103/physrevlett.85.3761} {\bibfield
  {journal} {\bibinfo  {journal} {Physical Review Letters}\ }\textbf {\bibinfo
  {volume} {85}},\ \bibinfo {pages} {3761–3764} (\bibinfo {year}
  {2000})}\BibitemShut {NoStop}%
\bibitem [{\citenamefont {Caprini}\ \emph {et~al.}(2016)\citenamefont {Caprini}
  \emph {et~al.}}]{Caprini:2015zlo}%
  \BibitemOpen
  \bibfield  {author} {\bibinfo {author} {\bibfnamefont {C.}~\bibnamefont
  {Caprini}} \emph {et~al.},\ }\href {\doibase 10.1088/1475-7516/2016/04/001}
  {\bibfield  {journal} {\bibinfo  {journal} {JCAP}\ }\textbf {\bibinfo
  {volume} {04}},\ \bibinfo {pages} {001} (\bibinfo {year} {2016})},\ \Eprint
  {http://arxiv.org/abs/1512.06239} {arXiv:1512.06239 [astro-ph.CO]}
  \BibitemShut {NoStop}%
\bibitem [{\citenamefont {Hindmarsh}\ \emph {et~al.}(2020)\citenamefont
  {Hindmarsh}, \citenamefont {L\"uben}, \citenamefont {Lumma},\ and\
  \citenamefont {Pauly}}]{Hindmarsh:2020hop}%
  \BibitemOpen
  \bibfield  {author} {\bibinfo {author} {\bibfnamefont {M.~B.}\ \bibnamefont
  {Hindmarsh}}, \bibinfo {author} {\bibfnamefont {M.}~\bibnamefont {L\"uben}},
  \bibinfo {author} {\bibfnamefont {J.}~\bibnamefont {Lumma}}, \ and\ \bibinfo
  {author} {\bibfnamefont {M.}~\bibnamefont {Pauly}},\ }\href@noop {} {\
  (\bibinfo {year} {2020})},\ \Eprint {http://arxiv.org/abs/2008.09136}
  {arXiv:2008.09136 [astro-ph.CO]} \BibitemShut {NoStop}%
\bibitem [{\citenamefont {Gasperini}\ and\ \citenamefont
  {Veneziano}(1993)}]{altCos_GW}%
  \BibitemOpen
  \bibfield  {author} {\bibinfo {author} {\bibfnamefont {M.}~\bibnamefont
  {Gasperini}}\ and\ \bibinfo {author} {\bibfnamefont {G.}~\bibnamefont
  {Veneziano}},\ }\href {\doibase 10.1016/0927-6505(93)90017-8} {\bibfield
  {journal} {\bibinfo  {journal} {Astroparticle Physics}\ }\textbf {\bibinfo
  {volume} {1}},\ \bibinfo {pages} {317–339} (\bibinfo {year}
  {1993})}\BibitemShut {NoStop}%
\bibitem [{\citenamefont {Gasperini}(2007)}]{altCos2_GW}%
  \BibitemOpen
  \bibfield  {author} {\bibinfo {author} {\bibfnamefont {M.}~\bibnamefont
  {Gasperini}},\ }\href {\doibase 10.1017/CBO9780511611285} {\emph {\bibinfo
  {title} {Elements of String Cosmology}}}\ (\bibinfo  {publisher} {Cambridge
  University Press},\ \bibinfo {year} {2007})\BibitemShut {NoStop}%
\bibitem [{\citenamefont {Aasi}\ \emph {et~al.}(2015)\citenamefont {Aasi} \emph
  {et~al.}}]{TheLIGOScientific:2014jea}%
  \BibitemOpen
  \bibfield  {author} {\bibinfo {author} {\bibfnamefont {J.}~\bibnamefont
  {Aasi}} \emph {et~al.} (\bibinfo {collaboration} {LIGO Scientific}),\ }\href
  {\doibase 10.1088/0264-9381/32/7/074001} {\bibfield  {journal} {\bibinfo
  {journal} {Class. Quant. Grav.}\ }\textbf {\bibinfo {volume} {32}},\ \bibinfo
  {pages} {074001} (\bibinfo {year} {2015})},\ \Eprint
  {http://arxiv.org/abs/1411.4547} {arXiv:1411.4547 [gr-qc]} \BibitemShut
  {NoStop}%
\bibitem [{\citenamefont {Acernese}\ \emph {et~al.}(2015)\citenamefont
  {Acernese} \emph {et~al.}}]{TheVirgo:2014hva}%
  \BibitemOpen
  \bibfield  {author} {\bibinfo {author} {\bibfnamefont {F.}~\bibnamefont
  {Acernese}} \emph {et~al.} (\bibinfo {collaboration} {VIRGO}),\ }\href
  {\doibase 10.1088/0264-9381/32/2/024001} {\bibfield  {journal} {\bibinfo
  {journal} {Class. Quant. Grav.}\ }\textbf {\bibinfo {volume} {32}},\ \bibinfo
  {pages} {024001} (\bibinfo {year} {2015})},\ \Eprint
  {http://arxiv.org/abs/1408.3978} {arXiv:1408.3978 [gr-qc]} \BibitemShut
  {NoStop}%
\bibitem [{\citenamefont {Abbott}\ \emph {et~al.}(2019)\citenamefont {Abbott}
  \emph {et~al.}}]{LIGOScientific:2019vic}%
  \BibitemOpen
  \bibfield  {author} {\bibinfo {author} {\bibfnamefont {B.~P.}\ \bibnamefont
  {Abbott}} \emph {et~al.} (\bibinfo {collaboration} {LIGO Scientific,
  Virgo}),\ }\href {\doibase 10.1103/PhysRevD.100.061101} {\bibfield  {journal}
  {\bibinfo  {journal} {Phys. Rev. D}\ }\textbf {\bibinfo {volume} {100}},\
  \bibinfo {pages} {061101} (\bibinfo {year} {2019})},\ \Eprint
  {http://arxiv.org/abs/1903.02886} {arXiv:1903.02886 [gr-qc]} \BibitemShut
  {NoStop}%
\bibitem [{\citenamefont {Abbott}\ \emph {et~al.}(2021)\citenamefont {Abbott}
  \emph {et~al.}}]{Abbott:2021xxi}%
  \BibitemOpen
  \bibfield  {author} {\bibinfo {author} {\bibfnamefont {R.}~\bibnamefont
  {Abbott}} \emph {et~al.} (\bibinfo {collaboration} {LIGO Scientific, Virgo,
  KAGRA}),\ }\href@noop {} {\  (\bibinfo {year} {2021})},\ \Eprint
  {http://arxiv.org/abs/2101.12130} {arXiv:2101.12130 [gr-qc]} \BibitemShut
  {NoStop}%
\bibitem [{\citenamefont {Abbott}\ \emph {et~al.}()\citenamefont {Abbott} \emph
  {et~al.}}]{O3Data}%
  \BibitemOpen
  \bibfield  {author} {\bibinfo {author} {\bibfnamefont {R.}~\bibnamefont
  {Abbott}} \emph {et~al.} (\bibinfo {collaboration} {LIGO Scientific
  Collaboration, Virgo Collaboration}),\ }\href@noop {} {}\bibinfo
  {howpublished} {\url{https://dcc.ligo.org/G2001287/public}}\BibitemShut
  {NoStop}%
\bibitem [{\citenamefont {Alexander}\ \emph {et~al.}(2006)\citenamefont
  {Alexander}, \citenamefont {Peskin},\ and\ \citenamefont
  {Sheikh-Jabbari}}]{Alexander_2006}%
  \BibitemOpen
  \bibfield  {author} {\bibinfo {author} {\bibfnamefont {S.~H.~S.}\
  \bibnamefont {Alexander}}, \bibinfo {author} {\bibfnamefont {M.~E.}\
  \bibnamefont {Peskin}}, \ and\ \bibinfo {author} {\bibfnamefont {M.~M.}\
  \bibnamefont {Sheikh-Jabbari}},\ }\href {\doibase
  10.1103/physrevlett.96.081301} {\bibfield  {journal} {\bibinfo  {journal}
  {Physical Review Letters}\ }\textbf {\bibinfo {volume} {96}} (\bibinfo {year}
  {2006}),\ 10.1103/physrevlett.96.081301}\BibitemShut {NoStop}%
\bibitem [{\citenamefont {Satoh}\ \emph {et~al.}(2008)\citenamefont {Satoh},
  \citenamefont {Kanno},\ and\ \citenamefont {Soda}}]{CS_PV}%
  \BibitemOpen
  \bibfield  {author} {\bibinfo {author} {\bibfnamefont {M.}~\bibnamefont
  {Satoh}}, \bibinfo {author} {\bibfnamefont {S.}~\bibnamefont {Kanno}}, \ and\
  \bibinfo {author} {\bibfnamefont {J.}~\bibnamefont {Soda}},\ }\href {\doibase
  10.1103/physrevd.77.023526} {\bibfield  {journal} {\bibinfo  {journal}
  {Physical Review D}\ }\textbf {\bibinfo {volume} {77}} (\bibinfo {year}
  {2008}),\ 10.1103/physrevd.77.023526}\BibitemShut {NoStop}%
\bibitem [{\citenamefont {Barnaby}\ and\ \citenamefont
  {Peloso}(2011)}]{AxInf_PV}%
  \BibitemOpen
  \bibfield  {author} {\bibinfo {author} {\bibfnamefont {N.}~\bibnamefont
  {Barnaby}}\ and\ \bibinfo {author} {\bibfnamefont {M.}~\bibnamefont
  {Peloso}},\ }\href {\doibase 10.1103/physrevlett.106.181301} {\bibfield
  {journal} {\bibinfo  {journal} {Physical Review Letters}\ }\textbf {\bibinfo
  {volume} {106}} (\bibinfo {year} {2011}),\
  10.1103/physrevlett.106.181301}\BibitemShut {NoStop}%
\bibitem [{\citenamefont {Kamionkowski}\ \emph {et~al.}(1994)\citenamefont
  {Kamionkowski}, \citenamefont {Kosowsky},\ and\ \citenamefont
  {Turner}}]{Kamionkowski_1994}%
  \BibitemOpen
  \bibfield  {author} {\bibinfo {author} {\bibfnamefont {M.}~\bibnamefont
  {Kamionkowski}}, \bibinfo {author} {\bibfnamefont {A.}~\bibnamefont
  {Kosowsky}}, \ and\ \bibinfo {author} {\bibfnamefont {M.~S.}\ \bibnamefont
  {Turner}},\ }\href {\doibase 10.1103/physrevd.49.2837} {\bibfield  {journal}
  {\bibinfo  {journal} {Physical Review D}\ }\textbf {\bibinfo {volume} {49}},\
  \bibinfo {pages} {2837–2851} (\bibinfo {year} {1994})}\BibitemShut
  {NoStop}%
\bibitem [{\citenamefont {Witten}(1984)}]{EdWitten_PhaseTrans}%
  \BibitemOpen
  \bibfield  {author} {\bibinfo {author} {\bibfnamefont {E.}~\bibnamefont
  {Witten}},\ }\href {\doibase 10.1103/PhysRevD.30.272} {\bibfield  {journal}
  {\bibinfo  {journal} {Phys. Rev. D}\ }\textbf {\bibinfo {volume} {30}},\
  \bibinfo {pages} {272} (\bibinfo {year} {1984})}\BibitemShut {NoStop}%
\bibitem [{\citenamefont {Hogan}(1986)}]{Hogan_PhaseTrans}%
  \BibitemOpen
  \bibfield  {author} {\bibinfo {author} {\bibfnamefont {C.~J.}\ \bibnamefont
  {Hogan}},\ }\href {\doibase 10.1093/mnras/218.4.629} {\bibfield  {journal}
  {\bibinfo  {journal} {MNRAS}\ }\textbf {\bibinfo {volume} {218}},\ \bibinfo
  {pages} {629} (\bibinfo {year} {1986})}\BibitemShut {NoStop}%
\bibitem [{\citenamefont {Brandenburg}\ \emph {et~al.}(1996)\citenamefont
  {Brandenburg}, \citenamefont {Enqvist},\ and\ \citenamefont
  {Olesen}}]{Brandenburg_1996}%
  \BibitemOpen
  \bibfield  {author} {\bibinfo {author} {\bibfnamefont {A.}~\bibnamefont
  {Brandenburg}}, \bibinfo {author} {\bibfnamefont {K.}~\bibnamefont
  {Enqvist}}, \ and\ \bibinfo {author} {\bibfnamefont {P.}~\bibnamefont
  {Olesen}},\ }\href {\doibase 10.1103/physrevd.54.1291} {\bibfield  {journal}
  {\bibinfo  {journal} {Physical Review D}\ }\textbf {\bibinfo {volume} {54}},\
  \bibinfo {pages} {1291–1300} (\bibinfo {year} {1996})}\BibitemShut
  {NoStop}%
\bibitem [{\citenamefont {Christensson}\ \emph {et~al.}(2001)\citenamefont
  {Christensson}, \citenamefont {Hindmarsh},\ and\ \citenamefont
  {Brandenburg}}]{Christensson_2001}%
  \BibitemOpen
  \bibfield  {author} {\bibinfo {author} {\bibfnamefont {M.}~\bibnamefont
  {Christensson}}, \bibinfo {author} {\bibfnamefont {M.}~\bibnamefont
  {Hindmarsh}}, \ and\ \bibinfo {author} {\bibfnamefont {A.}~\bibnamefont
  {Brandenburg}},\ }\href {\doibase 10.1103/physreve.64.056405} {\bibfield
  {journal} {\bibinfo  {journal} {Physical Review E}\ }\textbf {\bibinfo
  {volume} {64}} (\bibinfo {year} {2001}),\
  10.1103/physreve.64.056405}\BibitemShut {NoStop}%
\bibitem [{\citenamefont {Kahniashvili}\ \emph {et~al.}(2010)\citenamefont
  {Kahniashvili}, \citenamefont {Brandenburg}, \citenamefont {Tevzadze},\ and\
  \citenamefont {Ratra}}]{Kahniashvili_2010}%
  \BibitemOpen
  \bibfield  {author} {\bibinfo {author} {\bibfnamefont {T.}~\bibnamefont
  {Kahniashvili}}, \bibinfo {author} {\bibfnamefont {A.}~\bibnamefont
  {Brandenburg}}, \bibinfo {author} {\bibfnamefont {A.~G.}\ \bibnamefont
  {Tevzadze}}, \ and\ \bibinfo {author} {\bibfnamefont {B.}~\bibnamefont
  {Ratra}},\ }\href {\doibase 10.1103/physrevd.81.123002} {\bibfield  {journal}
  {\bibinfo  {journal} {Physical Review D}\ }\textbf {\bibinfo {volume} {81}}
  (\bibinfo {year} {2010}),\ 10.1103/physrevd.81.123002}\BibitemShut {NoStop}%
\bibitem [{\citenamefont {Brandenburg}\ \emph {et~al.}(2019)\citenamefont
  {Brandenburg}, \citenamefont {Kahniashvili}, \citenamefont {Mandal},
  \citenamefont {Pol}, \citenamefont {Tevzadze},\ and\ \citenamefont
  {Vachaspati}}]{Brandenburg_2019}%
  \BibitemOpen
  \bibfield  {author} {\bibinfo {author} {\bibfnamefont {A.}~\bibnamefont
  {Brandenburg}}, \bibinfo {author} {\bibfnamefont {T.}~\bibnamefont
  {Kahniashvili}}, \bibinfo {author} {\bibfnamefont {S.}~\bibnamefont
  {Mandal}}, \bibinfo {author} {\bibfnamefont {A.~R.}\ \bibnamefont {Pol}},
  \bibinfo {author} {\bibfnamefont {A.~G.}\ \bibnamefont {Tevzadze}}, \ and\
  \bibinfo {author} {\bibfnamefont {T.}~\bibnamefont {Vachaspati}},\ }\href
  {\doibase 10.1103/physrevfluids.4.024608} {\bibfield  {journal} {\bibinfo
  {journal} {Physical Review Fluids}\ }\textbf {\bibinfo {volume} {4}}
  (\bibinfo {year} {2019}),\ 10.1103/physrevfluids.4.024608}\BibitemShut
  {NoStop}%
\bibitem [{\citenamefont {Brandenburg}\ \emph {et~al.}(2021)\citenamefont
  {Brandenburg}, \citenamefont {He}, \citenamefont {Kahniashvili},
  \citenamefont {Rheinhardt},\ and\ \citenamefont
  {Schober}}]{Brandenburg:2021aln}%
  \BibitemOpen
  \bibfield  {author} {\bibinfo {author} {\bibfnamefont {A.}~\bibnamefont
  {Brandenburg}}, \bibinfo {author} {\bibfnamefont {Y.}~\bibnamefont {He}},
  \bibinfo {author} {\bibfnamefont {T.}~\bibnamefont {Kahniashvili}}, \bibinfo
  {author} {\bibfnamefont {M.}~\bibnamefont {Rheinhardt}}, \ and\ \bibinfo
  {author} {\bibfnamefont {J.}~\bibnamefont {Schober}},\ }\href@noop {} {\
  (\bibinfo {year} {2021})},\ \Eprint {http://arxiv.org/abs/2101.08178}
  {arXiv:2101.08178 [astro-ph.CO]} \BibitemShut {NoStop}%
\bibitem [{\citenamefont {Crowder}\ \emph {et~al.}(2013)\citenamefont
  {Crowder}, \citenamefont {Namba}, \citenamefont {Mandic}, \citenamefont
  {Mukohyama},\ and\ \citenamefont {Peloso}}]{Crowder_2013}%
  \BibitemOpen
  \bibfield  {author} {\bibinfo {author} {\bibfnamefont {S.}~\bibnamefont
  {Crowder}}, \bibinfo {author} {\bibfnamefont {R.}~\bibnamefont {Namba}},
  \bibinfo {author} {\bibfnamefont {V.}~\bibnamefont {Mandic}}, \bibinfo
  {author} {\bibfnamefont {S.}~\bibnamefont {Mukohyama}}, \ and\ \bibinfo
  {author} {\bibfnamefont {M.}~\bibnamefont {Peloso}},\ }\href {\doibase
  10.1016/j.physletb.2013.08.077} {\bibfield  {journal} {\bibinfo  {journal}
  {Physics Letters B}\ }\textbf {\bibinfo {volume} {726}},\ \bibinfo {pages}
  {66–71} (\bibinfo {year} {2013})}\BibitemShut {NoStop}%
\bibitem [{obs(2009)}]{obsRun_2009}%
  \BibitemOpen
  \href {\doibase 10.1038/nature08278} {\bibfield  {journal} {\bibinfo
  {journal} {Nature}\ }\textbf {\bibinfo {volume} {460}},\ \bibinfo {pages}
  {990–994} (\bibinfo {year} {2009})}\BibitemShut {NoStop}%
\bibitem [{\citenamefont {Lesieur}(1997)}]{HT_Study}%
  \BibitemOpen
  \bibfield  {author} {\bibinfo {author} {\bibfnamefont {M.}~\bibnamefont
  {Lesieur}},\ }\href@noop {} {\emph {\bibinfo {title} {Turbulence in
  Fluids}}}\ (\bibinfo  {publisher} {Springer Netherlands},\ \bibinfo {year}
  {1997})\BibitemShut {NoStop}%
\bibitem [{\citenamefont {Moiseev}\ and\ \citenamefont
  {Chkhetiani}(1996)}]{HK_Study}%
  \BibitemOpen
  \bibfield  {author} {\bibinfo {author} {\bibfnamefont {S.~S.}\ \bibnamefont
  {Moiseev}}\ and\ \bibinfo {author} {\bibfnamefont {O.}~\bibnamefont
  {Chkhetiani}},\ }\href@noop {} {\bibfield  {journal} {\bibinfo  {journal}
  {Journal of Experimental and Theoretical Physics}\ }\textbf {\bibinfo
  {volume} {83}},\ \bibinfo {pages} {192} (\bibinfo {year} {1996})}\BibitemShut
  {NoStop}%
\bibitem [{\citenamefont {Seto}\ and\ \citenamefont
  {Taruya}(2007)}]{Formalism_MathMod}%
  \BibitemOpen
  \bibfield  {author} {\bibinfo {author} {\bibfnamefont {N.}~\bibnamefont
  {Seto}}\ and\ \bibinfo {author} {\bibfnamefont {A.}~\bibnamefont {Taruya}},\
  }\href {\doibase 10.1103/PhysRevLett.99.121101} {\bibfield  {journal}
  {\bibinfo  {journal} {Phys. Rev. Lett.}\ }\textbf {\bibinfo {volume} {99}},\
  \bibinfo {pages} {121101} (\bibinfo {year} {2007})}\BibitemShut {NoStop}%
\bibitem [{\citenamefont {Akutsu}\ \emph {et~al.}(2020)\citenamefont {Akutsu}
  \emph {et~al.}}]{akutsu2020overview}%
  \BibitemOpen
  \bibfield  {author} {\bibinfo {author} {\bibfnamefont {T.}~\bibnamefont
  {Akutsu}} \emph {et~al.},\ }\href@noop {} {\enquote {\bibinfo {title}
  {Overview of kagra: Detector design and construction history},}\ } (\bibinfo
  {year} {2020}),\ \Eprint {http://arxiv.org/abs/2005.05574} {arXiv:2005.05574
  [physics.ins-det]} \BibitemShut {NoStop}%
\bibitem [{\citenamefont {Romano}\ and\ \citenamefont
  {Cornish}(2017)}]{Romano_Cornish}%
  \BibitemOpen
  \bibfield  {author} {\bibinfo {author} {\bibfnamefont {J.~D.}\ \bibnamefont
  {Romano}}\ and\ \bibinfo {author} {\bibfnamefont {N.~J.}\ \bibnamefont
  {Cornish}},\ }\href {\doibase 10.1007/s41114-017-0004-1} {\bibfield
  {journal} {\bibinfo  {journal} {Living Reviews in Relativity}\ }\textbf
  {\bibinfo {volume} {20}} (\bibinfo {year} {2017}),\
  10.1007/s41114-017-0004-1}\BibitemShut {NoStop}%
\bibitem [{\citenamefont {Matas}\ and\ \citenamefont
  {Romano}(2020)}]{Matas:2020roi}%
  \BibitemOpen
  \bibfield  {author} {\bibinfo {author} {\bibfnamefont {A.}~\bibnamefont
  {Matas}}\ and\ \bibinfo {author} {\bibfnamefont {J.~D.}\ \bibnamefont
  {Romano}},\ }\href@noop {} {\  (\bibinfo {year} {2020})},\ \Eprint
  {http://arxiv.org/abs/2012.00907} {arXiv:2012.00907 [gr-qc]} \BibitemShut
  {NoStop}%
\bibitem [{\citenamefont {Allen}\ and\ \citenamefont
  {Romano}(1999)}]{Formalism_Source}%
  \BibitemOpen
  \bibfield  {author} {\bibinfo {author} {\bibfnamefont {B.}~\bibnamefont
  {Allen}}\ and\ \bibinfo {author} {\bibfnamefont {J.~D.}\ \bibnamefont
  {Romano}},\ }\href {\doibase 10.1103/PhysRevD.59.102001} {\bibfield
  {journal} {\bibinfo  {journal} {Phys. Rev. D}\ }\textbf {\bibinfo {volume}
  {59}},\ \bibinfo {pages} {102001} (\bibinfo {year} {1999})}\BibitemShut
  {NoStop}%
\bibitem [{\citenamefont {Coughlin}\ \emph {et~al.}(2018)\citenamefont
  {Coughlin} \emph {et~al.}}]{Cirone:2018guh}%
  \BibitemOpen
  \bibfield  {author} {\bibinfo {author} {\bibfnamefont {M.~W.}\ \bibnamefont
  {Coughlin}} \emph {et~al.},\ }\href {\doibase 10.1103/PhysRevD.97.102007}
  {\bibfield  {journal} {\bibinfo  {journal} {Phys. Rev. D}\ }\textbf {\bibinfo
  {volume} {97}},\ \bibinfo {pages} {102007} (\bibinfo {year} {2018})},\
  \Eprint {http://arxiv.org/abs/1802.00885} {arXiv:1802.00885 [gr-qc]}
  \BibitemShut {NoStop}%
\bibitem [{\citenamefont {Meyers}\ \emph {et~al.}(2020)\citenamefont {Meyers},
  \citenamefont {Martinovic}, \citenamefont {Christensen},\ and\ \citenamefont
  {Sakellariadou}}]{Meyers:2020qrb}%
  \BibitemOpen
  \bibfield  {author} {\bibinfo {author} {\bibfnamefont {P.~M.}\ \bibnamefont
  {Meyers}}, \bibinfo {author} {\bibfnamefont {K.}~\bibnamefont {Martinovic}},
  \bibinfo {author} {\bibfnamefont {N.}~\bibnamefont {Christensen}}, \ and\
  \bibinfo {author} {\bibfnamefont {M.}~\bibnamefont {Sakellariadou}},\ }\href
  {\doibase 10.1103/PhysRevD.102.102005} {\bibfield  {journal} {\bibinfo
  {journal} {Phys. Rev. D}\ }\textbf {\bibinfo {volume} {102}},\ \bibinfo
  {pages} {102005} (\bibinfo {year} {2020})},\ \Eprint
  {http://arxiv.org/abs/2008.00789} {arXiv:2008.00789 [gr-qc]} \BibitemShut
  {NoStop}%
\bibitem [{\citenamefont {Long}\ \emph {et~al.}(2014)\citenamefont {Long},
  \citenamefont {Sabancilar},\ and\ \citenamefont {Vachaspati}}]{Long_2014}%
  \BibitemOpen
  \bibfield  {author} {\bibinfo {author} {\bibfnamefont {A.~J.}\ \bibnamefont
  {Long}}, \bibinfo {author} {\bibfnamefont {E.}~\bibnamefont {Sabancilar}}, \
  and\ \bibinfo {author} {\bibfnamefont {T.}~\bibnamefont {Vachaspati}},\
  }\href {\doibase 10.1088/1475-7516/2014/02/036} {\bibfield  {journal}
  {\bibinfo  {journal} {Journal of Cosmology and Astroparticle Physics}\
  }\textbf {\bibinfo {volume} {2014}},\ \bibinfo {pages} {036–036} (\bibinfo
  {year} {2014})}\BibitemShut {NoStop}%
\bibitem [{\citenamefont {Dorsch}\ \emph {et~al.}(2017)\citenamefont {Dorsch},
  \citenamefont {Huber}, \citenamefont {Konstandin},\ and\ \citenamefont
  {No}}]{Dorsch_2017}%
  \BibitemOpen
  \bibfield  {author} {\bibinfo {author} {\bibfnamefont {G.}~\bibnamefont
  {Dorsch}}, \bibinfo {author} {\bibfnamefont {S.}~\bibnamefont {Huber}},
  \bibinfo {author} {\bibfnamefont {T.}~\bibnamefont {Konstandin}}, \ and\
  \bibinfo {author} {\bibfnamefont {J.}~\bibnamefont {No}},\ }\href {\doibase
  10.1088/1475-7516/2017/05/052} {\bibfield  {journal} {\bibinfo  {journal}
  {Journal of Cosmology and Astroparticle Physics}\ }\textbf {\bibinfo {volume}
  {2017}},\ \bibinfo {pages} {052–052} (\bibinfo {year} {2017})}\BibitemShut
  {NoStop}%
\bibitem [{\citenamefont {Kahniashvili}(2005)}]{Tina_2005}%
  \BibitemOpen
  \bibfield  {author} {\bibinfo {author} {\bibfnamefont {T.}~\bibnamefont
  {Kahniashvili}},\ }\href@noop {} {\  (\bibinfo {year} {2005})},\ \Eprint
  {http://arxiv.org/abs/astro-ph/0508459} {arXiv:astro-ph/0508459} \BibitemShut
  {NoStop}%
\bibitem [{\citenamefont {Roper~Pol}\ \emph {et~al.}(2020)\citenamefont
  {Roper~Pol}, \citenamefont {Mandal}, \citenamefont {Brandenburg},
  \citenamefont {Kahniashvili},\ and\ \citenamefont {Kosowsky}}]{Pol:2019yex}%
  \BibitemOpen
  \bibfield  {author} {\bibinfo {author} {\bibfnamefont {A.}~\bibnamefont
  {Roper~Pol}}, \bibinfo {author} {\bibfnamefont {S.}~\bibnamefont {Mandal}},
  \bibinfo {author} {\bibfnamefont {A.}~\bibnamefont {Brandenburg}}, \bibinfo
  {author} {\bibfnamefont {T.}~\bibnamefont {Kahniashvili}}, \ and\ \bibinfo
  {author} {\bibfnamefont {A.}~\bibnamefont {Kosowsky}},\ }\href {\doibase
  10.1103/PhysRevD.102.083512} {\bibfield  {journal} {\bibinfo  {journal}
  {Phys. Rev. D}\ }\textbf {\bibinfo {volume} {102}},\ \bibinfo {pages}
  {083512} (\bibinfo {year} {2020})},\ \Eprint
  {http://arxiv.org/abs/1903.08585} {arXiv:1903.08585 [astro-ph.CO]}
  \BibitemShut {NoStop}%
\bibitem [{\citenamefont {Weir}(2018)}]{Weir_2018}%
  \BibitemOpen
  \bibfield  {author} {\bibinfo {author} {\bibfnamefont {D.~J.}\ \bibnamefont
  {Weir}},\ }\href {\doibase 10.1098/rsta.2017.0126} {\bibfield  {journal}
  {\bibinfo  {journal} {Philosophical Transactions of the Royal Society A:
  Mathematical, Physical and Engineering Sciences}\ }\textbf {\bibinfo {volume}
  {376}},\ \bibinfo {pages} {20170126} (\bibinfo {year} {2018})}\BibitemShut
  {NoStop}%
\bibitem [{\citenamefont {Kosowsky}\ \emph {et~al.}(2002)\citenamefont
  {Kosowsky}, \citenamefont {Mack},\ and\ \citenamefont
  {Kahniashvili}}]{Tina_2002}%
  \BibitemOpen
  \bibfield  {author} {\bibinfo {author} {\bibfnamefont {A.}~\bibnamefont
  {Kosowsky}}, \bibinfo {author} {\bibfnamefont {A.}~\bibnamefont {Mack}}, \
  and\ \bibinfo {author} {\bibfnamefont {T.}~\bibnamefont {Kahniashvili}},\
  }\href {\doibase 10.1103/physrevd.66.024030} {\bibfield  {journal} {\bibinfo
  {journal} {Physical Review D}\ }\textbf {\bibinfo {volume} {66}} (\bibinfo
  {year} {2002}),\ 10.1103/physrevd.66.024030}\BibitemShut {NoStop}%
\bibitem [{\citenamefont {Kahniashvili}\ \emph {et~al.}(2020)\citenamefont
  {Kahniashvili}, \citenamefont {Brandenburg}, \citenamefont {Gogoberidze},
  \citenamefont {Mandal},\ and\ \citenamefont
  {Pol}}]{kahniashvili2020circular}%
  \BibitemOpen
  \bibfield  {author} {\bibinfo {author} {\bibfnamefont {T.}~\bibnamefont
  {Kahniashvili}}, \bibinfo {author} {\bibfnamefont {A.}~\bibnamefont
  {Brandenburg}}, \bibinfo {author} {\bibfnamefont {G.}~\bibnamefont
  {Gogoberidze}}, \bibinfo {author} {\bibfnamefont {S.}~\bibnamefont {Mandal}},
  \ and\ \bibinfo {author} {\bibfnamefont {A.~R.}\ \bibnamefont {Pol}},\
  }\href@noop {} {\  (\bibinfo {year} {2020})},\ \Eprint
  {http://arxiv.org/abs/2011.05556} {arXiv:2011.05556 [astro-ph.CO]}
  \BibitemShut {NoStop}%
\bibitem [{\citenamefont {Abbott}\ \emph {et~al.}(2018)\citenamefont {Abbott}
  \emph {et~al.}}]{Aasi:2013wya}%
  \BibitemOpen
  \bibfield  {author} {\bibinfo {author} {\bibfnamefont {B.~P.}\ \bibnamefont
  {Abbott}} \emph {et~al.} (\bibinfo {collaboration} {KAGRA, LIGO Scientific,
  VIRGO}),\ }\href {\doibase 10.1007/s41114-018-0012-9} {\bibfield  {journal}
  {\bibinfo  {journal} {Living Rev. Rel.}\ }\textbf {\bibinfo {volume} {21}},\
  \bibinfo {pages} {3} (\bibinfo {year} {2018})},\ \Eprint
  {http://arxiv.org/abs/1304.0670} {arXiv:1304.0670 [gr-qc]} \BibitemShut
  {NoStop}%
\bibitem [{\citenamefont {Punturo}\ \emph {et~al.}(2010)\citenamefont {Punturo}
  \emph {et~al.}}]{Punturo_2010}%
  \BibitemOpen
  \bibfield  {author} {\bibinfo {author} {\bibfnamefont {M.}~\bibnamefont
  {Punturo}} \emph {et~al.},\ }\href {\doibase 10.1088/0264-9381/27/8/084007}
  {\bibfield  {journal} {\bibinfo  {journal} {Classical and Quantum Gravity}\
  }\textbf {\bibinfo {volume} {27}},\ \bibinfo {pages} {084007} (\bibinfo
  {year} {2010})}\BibitemShut {NoStop}%
\bibitem [{\citenamefont {Reitze}\ \emph {et~al.}(2019)\citenamefont {Reitze},
  \citenamefont {Adhikari}, \citenamefont {Ballmer}, \citenamefont {Barish},
  \citenamefont {Barsotti}, \citenamefont {Billingsley}, \citenamefont {Brown},
  \citenamefont {Chen}, \citenamefont {Coyne}, \citenamefont {Eisenstein},
  \citenamefont {Evans}, \citenamefont {Fritschel}, \citenamefont {Hall},
  \citenamefont {Lazzarini}, \citenamefont {Lovelace}, \citenamefont {Read},
  \citenamefont {Sathyaprakash}, \citenamefont {Shoemaker}, \citenamefont
  {Smith}, \citenamefont {Torrie}, \citenamefont {Vitale}, \citenamefont
  {Weiss}, \citenamefont {Wipf},\ and\ \citenamefont
  {Zucker}}]{reitze2019cosmic}%
  \BibitemOpen
  \bibfield  {author} {\bibinfo {author} {\bibfnamefont {D.}~\bibnamefont
  {Reitze}}, \bibinfo {author} {\bibfnamefont {R.~X.}\ \bibnamefont
  {Adhikari}}, \bibinfo {author} {\bibfnamefont {S.}~\bibnamefont {Ballmer}},
  \bibinfo {author} {\bibfnamefont {B.}~\bibnamefont {Barish}}, \bibinfo
  {author} {\bibfnamefont {L.}~\bibnamefont {Barsotti}}, \bibinfo {author}
  {\bibfnamefont {G.}~\bibnamefont {Billingsley}}, \bibinfo {author}
  {\bibfnamefont {D.~A.}\ \bibnamefont {Brown}}, \bibinfo {author}
  {\bibfnamefont {Y.}~\bibnamefont {Chen}}, \bibinfo {author} {\bibfnamefont
  {D.}~\bibnamefont {Coyne}}, \bibinfo {author} {\bibfnamefont
  {R.}~\bibnamefont {Eisenstein}}, \bibinfo {author} {\bibfnamefont
  {M.}~\bibnamefont {Evans}}, \bibinfo {author} {\bibfnamefont
  {P.}~\bibnamefont {Fritschel}}, \bibinfo {author} {\bibfnamefont {E.~D.}\
  \bibnamefont {Hall}}, \bibinfo {author} {\bibfnamefont {A.}~\bibnamefont
  {Lazzarini}}, \bibinfo {author} {\bibfnamefont {G.}~\bibnamefont {Lovelace}},
  \bibinfo {author} {\bibfnamefont {J.}~\bibnamefont {Read}}, \bibinfo {author}
  {\bibfnamefont {B.~S.}\ \bibnamefont {Sathyaprakash}}, \bibinfo {author}
  {\bibfnamefont {D.}~\bibnamefont {Shoemaker}}, \bibinfo {author}
  {\bibfnamefont {J.}~\bibnamefont {Smith}}, \bibinfo {author} {\bibfnamefont
  {C.}~\bibnamefont {Torrie}}, \bibinfo {author} {\bibfnamefont
  {S.}~\bibnamefont {Vitale}}, \bibinfo {author} {\bibfnamefont
  {R.}~\bibnamefont {Weiss}}, \bibinfo {author} {\bibfnamefont
  {C.}~\bibnamefont {Wipf}}, \ and\ \bibinfo {author} {\bibfnamefont
  {M.}~\bibnamefont {Zucker}},\ }\href@noop {} {\enquote {\bibinfo {title}
  {Cosmic explorer: The u.s. contribution to gravitational-wave astronomy
  beyond ligo},}\ } (\bibinfo {year} {2019}),\ \Eprint
  {http://arxiv.org/abs/1907.04833} {arXiv:1907.04833 [astro-ph.IM]}
  \BibitemShut {NoStop}%
\bibitem [{\citenamefont {{LISA collaboration}}(2017)}]{LISA_Prop}%
  \BibitemOpen
  \bibfield  {author} {\bibinfo {author} {\bibnamefont {{LISA
  collaboration}}},\ }\href@noop {} {\enquote {\bibinfo {title} {Laser
  interferometer space antenna},}\ } (\bibinfo {year} {2017}),\ \Eprint
  {http://arxiv.org/abs/1702.00786} {arXiv:1702.00786 [astro-ph.IM]}
  \BibitemShut {NoStop}%
\bibitem [{\citenamefont {Sorbo}(2011)}]{Sorbo_2011}%
  \BibitemOpen
  \bibfield  {author} {\bibinfo {author} {\bibfnamefont {L.}~\bibnamefont
  {Sorbo}},\ }\href {\doibase 10.1088/1475-7516/2011/06/003} {\bibfield
  {journal} {\bibinfo  {journal} {Journal of Cosmology and Astroparticle
  Physics}\ }\textbf {\bibinfo {volume} {2011}},\ \bibinfo {pages} {003–003}
  (\bibinfo {year} {2011})}\BibitemShut {NoStop}%
\bibitem [{\citenamefont {Hunter}(2007)}]{Hunter:2007}%
  \BibitemOpen
  \bibfield  {author} {\bibinfo {author} {\bibfnamefont {J.~D.}\ \bibnamefont
  {Hunter}},\ }\href {\doibase 10.1109/MCSE.2007.55} {\bibfield  {journal}
  {\bibinfo  {journal} {Computing in Science \& Engineering}\ }\textbf
  {\bibinfo {volume} {9}},\ \bibinfo {pages} {90} (\bibinfo {year}
  {2007})}\BibitemShut {NoStop}%
\bibitem [{\citenamefont {{van der Walt}}\ \emph {et~al.}(2011)\citenamefont
  {{van der Walt}}, \citenamefont {{Colbert}},\ and\ \citenamefont
  {{Varoquaux}}}]{numpy}%
  \BibitemOpen
  \bibfield  {author} {\bibinfo {author} {\bibfnamefont {S.}~\bibnamefont {{van
  der Walt}}}, \bibinfo {author} {\bibfnamefont {S.~C.}\ \bibnamefont
  {{Colbert}}}, \ and\ \bibinfo {author} {\bibfnamefont {G.}~\bibnamefont
  {{Varoquaux}}},\ }\href@noop {} {\bibfield  {journal} {\bibinfo  {journal}
  {Computing in Science Engineering}\ }\textbf {\bibinfo {volume} {13}},\
  \bibinfo {pages} {22} (\bibinfo {year} {2011})}\BibitemShut {NoStop}%
\bibitem [{\citenamefont {Ashton}\ \emph {et~al.}(2019)\citenamefont {Ashton}
  \emph {et~al.}}]{Ashton:2018jfp}%
  \BibitemOpen
  \bibfield  {author} {\bibinfo {author} {\bibfnamefont {G.}~\bibnamefont
  {Ashton}} \emph {et~al.},\ }\href {\doibase 10.3847/1538-4365/ab06fc}
  {\bibfield  {journal} {\bibinfo  {journal} {Astrophys. J. Suppl.}\ }\textbf
  {\bibinfo {volume} {241}},\ \bibinfo {pages} {27} (\bibinfo {year} {2019})},\
  \Eprint {http://arxiv.org/abs/1811.02042} {arXiv:1811.02042 [astro-ph.IM]}
  \BibitemShut {NoStop}%
\bibitem [{\citenamefont {{Hinton}}(2016)}]{Hinton2016}%
  \BibitemOpen
  \bibfield  {author} {\bibinfo {author} {\bibfnamefont {S.~R.}\ \bibnamefont
  {{Hinton}}},\ }\href {\doibase 10.21105/joss.00045} {\bibfield  {journal}
  {\bibinfo  {journal} {The Journal of Open Source Software}\ }\textbf
  {\bibinfo {volume} {1}},\ \bibinfo {eid} {00045} (\bibinfo {year}
  {2016})}\BibitemShut {NoStop}%
\end{thebibliography}%

\appendix



\end{document}